\documentclass[twocolumn,showpacs,superscriptaddress,10pt]{revtex4-2}
\usepackage{amsmath,amssymb,graphicx, subfigure, hyperref, braket, float}
\usepackage{blindtext}

\counterwithout{figure}{section}
\usepackage[export]{adjustbox}
\graphicspath{{figure/}}
\usepackage{graphicx, xcolor}
\usepackage[normalem]{ulem}
\usepackage[margin=.8in,text={10in,10.5in},centering]{geometry}
\usepackage{graphicx}
\usepackage{hyperref}
\usepackage{lipsum}
\usepackage{caption}
\usepackage{tikz}
\usepackage{needspace} 
\usepackage{bbold}
\usepackage{pgf}
\usepackage{ragged2e} 

\definecolor{editorialgreen}{RGB}{0,128,96}

\allowdisplaybreaks
\hypersetup{ colorlinks=true,linkcolor=red,citecolor=blue,urlcolor=magenta}
\begin{document}
	
\title{Phase-switchable nonreciprocal entanglement via magnon squeezing in ring-cavity optomagnomechanics}

\author{Z. Imara} \email{imara.ziyad@etu.uae.ac.ma}
\address{Laboratory of R\&D in Engineering Sciences, Faculty of Sciences and Techniques Al-Hoceima, Abdelmalek Essaadi University, Tetouan, Morocco}
\address{The UM6P Vanguard Center, Mohammed VI Polytechnic University (UM6P), Rocade Rabat-Salé, Technopolis, 11103 Morocco}

\author{A. El Allati}\email{eabderrahim@uae.ac.ma}
\address{Laboratory of R\&D in Engineering Sciences, Faculty of Sciences and Techniques Al-Hoceima, Abdelmalek Essaadi University, Tetouan, Morocco}
\address{Université Grenoble Alpes, CNRS, LPMMC, 38000 Grenoble, France.}

\author{A. Belfakir}\email{Abdessamad.BELFAKIR@um6p.ma}
\address{The UM6P Vanguard Center, Mohammed VI Polytechnic University (UM6P), Rocade Rabat-Salé, Technopolis, 11103 Morocco}

\author{K. El Anouz}\email{kelanouz@uae.ac.ma}
\address{Laboratory of R\&D in Engineering Sciences, Faculty of Sciences and Techniques Al-Hoceima, Abdelmalek Essaadi University, Tetouan, Morocco}

\author{I. P\'erez Castillo} \email{iperez@izt.uam.mx}
\address{Departamento de Física, Universidad Autónoma Metropolitana-Iztapalapa, San Rafael Atlixco 186, Ciudad de México 09340, Mexico}

	\begin{abstract} 
Cavity optomagnomechanics provides a versatile platform to explore macroscopic quantum correlations, particularly nonreciprocal entanglement. In this work, we propose a theoretical scheme to generate \emph{switchable} bipartite and tripartite entanglement in an optomagnomechanical ring cavity by exploiting phase-controlled magnon squeezing. Indeed, two spatially separated ferrimagnetic YIG microbridges become entangled through their magnetostriction-mediated coupling to mechanical motion and a common cavity field via radiation-pressure interaction. The squeezing process introduces two phase-dependent contributions to the magnon response, namely an effective detuning shift {$\Delta_{\theta_j}$} and a quadrature-damping contribution {$\kappa_{\theta_j}$}, both of which reverse sign upon a $\pi$ phase shift, providing an \textit{in situ} control to switch the entanglement response. The nonreciprocal entanglement is defined operationally through the asymmetric entanglement response under the phase reversal $\theta_j \to \theta_j + \pi$, quantified by normalized contrast ratios $C_E$ and $C_{\mathcal{R}}$, which measure the relative difference between the entanglement obtained at $\theta_j$ and at the phase-reversed configuration $\theta_j+\pi$. The resulting phase-tuning method provides a flexible and robust route to achieve high-contrast bipartite and tripartite entanglement within stable parameter regions, establishing magnon squeezing as a practical quantum resource for switchable quantum correlations in hybrid platforms.
	\end{abstract}
	
	\vspace{2cm}
	\maketitle

	\section{Introduction}
	\label{SECI}
	
Recently, the control and the manipulation of macroscopic quantum states have become a central theme in physics, bridging fundamental science and emerging quantum technologies \cite{A1}. In this context, cavity optomechanics (OM) has established itself as a leading platform, enabling phenomena such as ground-state cooling, squeezing, and multipartite entanglement via radiation-pressure interactions between optical and mechanical modes \cite{A2,A3,A4,A5}. Moreover, cavity magnomechanics (MM) has emerged as a complementary and equally promising framework \cite{A6,A66,A666}. For these hybrid systems, magnons, which are collective spin excitations in ferrimagnetic crystals such as yttrium iron garnet (YIG, Y$_3$Fe$_5$O$_{12}$), can be coupled to mechanical vibrations through magnetostrictive forces. This interaction has enabled a wide range of nonclassical effects, including nonreciprocal magnon-induced entanglement \cite{A7}, magnon-photon-phonon entanglement \cite{A8}, and magnon blockade \cite{A9}.

The continued development of OM and MM platforms has naturally led to optomagnomechanical (OMM) cavities, which combine radiation-pressure and magnetostrictive interactions in a single hybrid setting \cite{A6}. These systems offer tunable couplings and nonlinear dynamics, providing a versatile way to control light, motion, and collective spin excitations. More generally, cavity magnonics has enabled interfaces between magnons and other quantum platforms, including Bose-Einstein condensates \cite{B0}, and quantum sensing applications such as cavity-based magnetometry \cite{add2}, opening additional opportunities for hybrid quantum networks \cite{B1}.

The magnon squeezing process provides a powerful resource for enhancing the performance of magnonic platforms, enabling improvements in ground-state cooling, amplification of nonlinear responses and the generation of strong quantum correlations \cite{B5,B6}. Indeed, several mechanisms have been proposed to generate squeezed magnon states, including intrinsic magnetostrictive nonlinearities \cite{B7}, reservoir-engineered cavity magnomechanics \cite{B8}, two-tone microwave driving \cite{B9}, and other related approaches \cite{C0,C1,C2}. Fan \textit{et al.} proposed an alternative route to squeeze magnons in optomagnomechanical systems \cite{C3}, and a recent study has explored entanglement generation via microwave-field squeezing in such systems \cite{D2}, further illustrating the flexibility of photon-magnon-phonon hybridization.

Nonreciprocal entanglement, characterized by an asymmetric entanglement response when a control parameter of the system is reversed \cite{C4,C9}, has been realized by breaking time-reversal symmetry through several mechanisms. The concept was first introduced in rotating optomechanical resonators, where mechanical rotation induces opposite frequency shifts for clockwise and counterclockwise propagating modes. For example, Jiao \textit{et al.} demonstrated controllable directionality of photon-phonon entanglement by inducing mode-dependent refractive indices \cite{C4,C44}. Analogous effects have since been explored in MM systems. Chen \textit{et al.} \cite{C9} exploited the Kerr nonlinearity, which induces a magnon frequency shift and an additional two-magnon effect. Lu \textit{et al.} \cite{D1} demonstrated nonreciprocity via the Barnett effect, where the induced frequency shift can be controlled to be positive or negative by adjusting the magnetic field direction. Other approaches include chiral couplings \cite{D0}, spinning the magnomechanical cavity \cite{C8}, among others \cite{C6,C5,D11,D111,C88,C888}. Many previous approaches rely primarily on effective frequency shifts, for example rotation-, Kerr-, or Barnett-induced shifts, whereas chiral-coupling schemes exploit coupling asymmetry. Magnon squeezing provides a complementary control mechanism because its phase modifies both an effective frequency contribution and a quadrature-damping contribution of the magnon mode. Recently, magnon squeezing has been proposed as a distinct mechanism for nonreciprocal entanglement in cavity magnomechanics \cite{C6}, introducing dual control where the squeezing phase simultaneously manipulates both the frequency shift and the dissipation rate.

In this paper, we employ this magnon-squeezing mechanism within an OMM ring cavity, where radiation-pressure and magnetostrictive interactions coexist. In our scheme, reversing the squeezing phase under $\theta \to \theta + \pi$ reverses both the effective frequency shift and the damping rate, producing two experimentally distinct configurations analogous to magnetic axis reversal in the Kerr effect \cite{C9} and rotation reversal in the Barnett effect \cite{D1}. Here, $\theta$ represents the phase of the parametric drive that generates the magnon squeezing, following the standard formalism of parametric interactions established by Agarwal and Huang \cite{add1}.
While Ref.~\cite{C6} demonstrated this dual control in a standard cavity magnomechanics with magnetostrictive coupling alone, our ring-cavity OMM architecture incorporates two spatially separated YIG microbridges coupled to a common optical cavity mode through both radiation-pressure and magnetostrictive interactions, enabling switchable nonreciprocal entanglement by selecting which magnon is squeezed. 
This channel-selection operation is not available in the single-magnon configuration considered in Ref.~\cite{C6}, where the nonreciprocal response is controlled by the squeezing phase of one magnon mode only. It therefore provides an additional degree of reconfigurability for distributing entanglement among spatially separated optomagnonic channels, rather than merely increasing the absolute value of the entanglement.
The coexistence of both interactions offers greater tunability through the interplay between OM and MM couplings. 
This \textit{in situ} phase control provides high-contrast nonreciprocal response in selected stable regions without requiring structural modification or mechanical rotation. We show how the squeezing amplitude, squeezing phase, bath temperature, and MM coupling enable robust nonreciprocal entanglement across these stable windows, and we verify the presence of genuine magnon squeezing for all configurations used in the study.

The paper is organized as follows. Sec.~\ref{sec:model} presents the optomagnomechanical ring-cavity Hamiltonian and derives the linearized quantum Langevin equations. Sec.~\ref{sec:results} investigates bipartite and tripartite entanglement under two squeezing scenarios: squeezing either the first or second magnon, demonstrating switchable entanglement with experimentally accessible parameters. Sec.~\ref{sec:conclusion} summarizes our main results and conclusions.

\section{THEORETICAL MODEL}
\label{sec:model}

\begin{figure}[b]	
	\begin{flushleft}
		\subfigure{\includegraphics[scale=0.4]{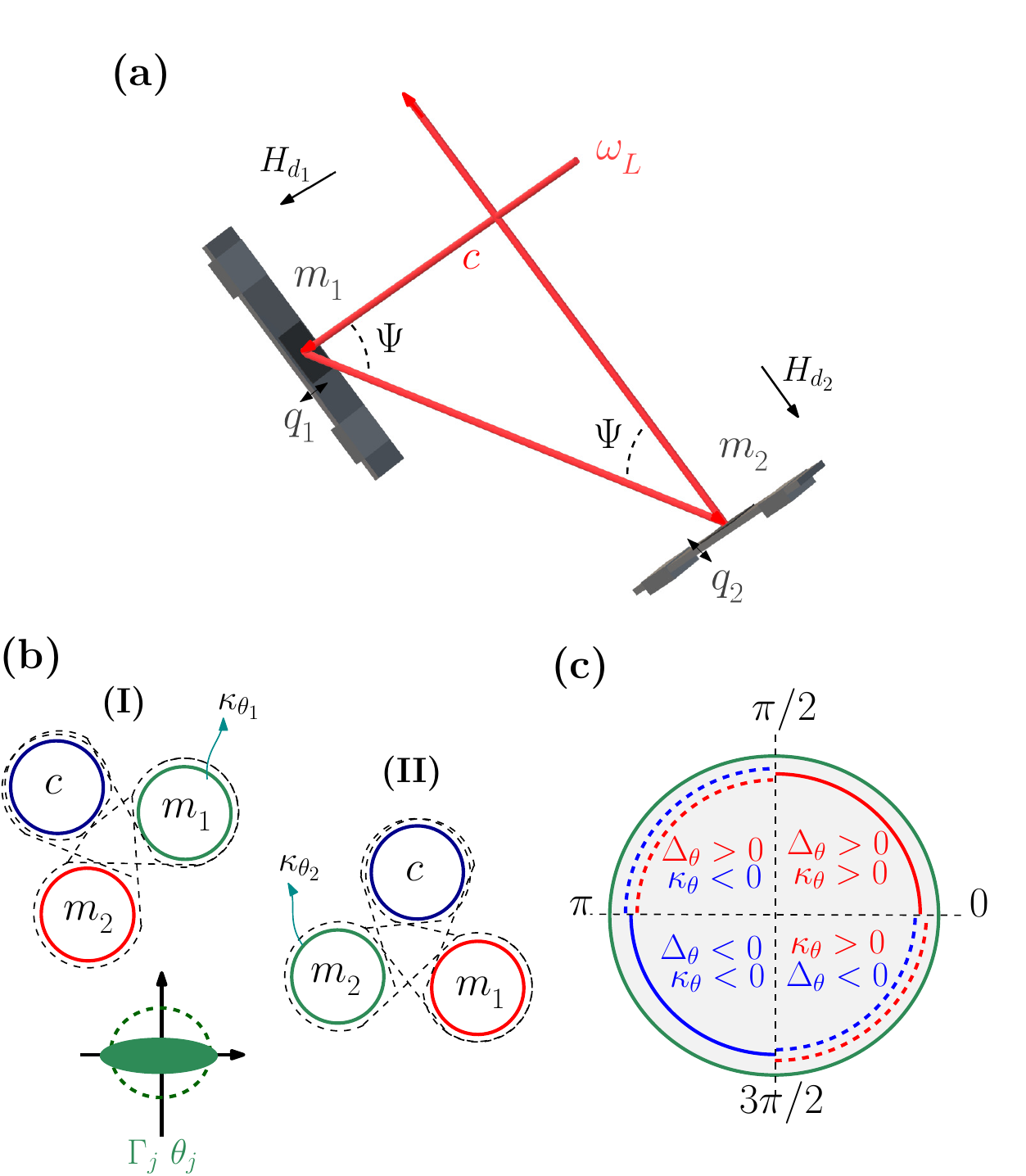}}
	\end{flushleft}
	\vspace{-1em}
	\captionsetup{justification=RaggedRight, singlelinecheck=false}
	\caption{{(a) Schematic of the optomagnomechanical ring-cavity. (b) Sketch of optomagnonic \emph{bipartite} entanglement for the two magnon-squeezing scenarios considered in this work, labeled (I)-(II); green modes indicate the squeezed magnon modes with amplitude $\Gamma_{j}$ and phase $\theta_j$. (c) Simplified schematic illustrating how changing the squeezing-phase sign modifies the effective magnon detuning and damping rate, enabling phase-controlled dissipation engineering.}}
	\label{fig:1}
\end{figure}

Let us consider a ring-cavity optomagnomechanics \cite{D3}, as illustrated in Fig.~\ref{fig:1}. The system consists of an optical cavity mode coupled to a magnonic (Kittel) mode \cite{D4} hosted in a YIG resonator. The mechanical vibrations of the YIG microbridge interact with the magnons via magnetostrictive deformation of the micron-sized YIG structure \cite{D04}. On the optomechanical side, the cavity is formed by a fixed mirror and a highly reflective micron-sized movable mirror attached to the YIG microbridge \cite{D5}. Owing to its small mass and dimensions, the movable mirror preserves the intrinsic mechanical properties of the YIG resonator.

For modeling purposes, we assume a uniform deformation mode perpendicular to the fixed surfaces and we neglect bending-related displacements. This geometry enables tight integration of the YIG microbridges with the optical elements \cite{B4,D6}. A static bias magnetic field is applied to each YIG bridge to generate the magnetostrictive interaction. The magnon mode is coherently driven by a microwave field at frequency $\omega_{0}$ (e.g., via a loop antenna). While, the optical cavity mode is driven by an external laser at frequency $\omega_{L}$ and coupled to the mechanical motion through radiation pressure. Hence, the Hamiltonian of this system in the rotating frame at $\omega_{0}$ and $\omega_{L}$ frequencies (setting $\hbar=1$) reads
\begin{eqnarray}
	\label{1}
	\mathbf{H}&=&\Delta_c c^\dagger c +\sum_{j=1,2}\Delta_{m_j} m_j^\dagger m_j+\frac{\omega_{b_j}}{2}\big({q_j^2+p_j^2}\big)\nonumber\\
	&+&\sum_{j=1,2}  \bar{g}_{0j}c^\dagger c\,{q_j}+g_{m_j}m_j^\dagger m_j\,{q_j} + i\big[\big( \frac{\Gamma_j}{2}e^{i\theta_j}m^{\dagger2}_j\nonumber
	\\&+&\Omega_j m_j^\dagger \big)-\text{h.c.}\big] +(\Upsilon c^\dagger -\text{h.c.}).
\end{eqnarray}
Here $c$ ($c^\dagger$) and $m_j$ ($m_j^\dagger$) are the annihilation (creation) operators of the optical cavity and the $j^{\text{th}}$ magnon mode, satisfying $[c,c^\dagger]=1$ and $[m_j,m_k^\dagger]=\delta_{jk}$. Moreover, ${\Delta}_{c}=\omega_{c}-\omega_L$, ${\Delta}_{m_j}=\omega_{m_j}-\omega_0$.
For the mechanical modes, $ q_j$ and $p_j$ are the dimensionless position and momentum quadratures of the $j$-th phonon mode, satisfying $[q_j,p_k]=i\delta_{jk}$.
The parameters $\omega_c$ and $\omega_{b_j}$ denote the optical and $j^{\text{th}}$ mechanical-mode frequencies, respectively. Moreover, each YIG microbridge is biased by a uniform magnetic field $H_{0j}$, which sets the magnon frequency $\omega_{m_j}=\gamma H_{0j}$, with $\gamma/2\pi=28\mathrm{GHz/T}$ \cite{A8}. A microwave drive of amplitude $H_{d_j}$ is applied perpendicular to $H_{0j}$, which excites the magnon mode with (Rabi) drive strength $\Omega_j=\sqrt{5}/4\gamma\sqrt{N_0}H_{d_j}$, where $N_0=\rho V$ is the number of spins, $\rho=4.22\times 10^{27} \mathrm{m}^{-3}$ is the spin density, and $V=5\times 2\times 1~\mu\mathrm{m}^3$ is the bridge volume, assuming the low-excitation regime $\langle m_j^\dagger m_j\rangle \ll 2N_0 s$ with $s=5/2$ for Fe$^{3+}$ in YIG. 
The term in Eq.~(\ref{1}), $[(\Gamma_j/2) e^{i\theta_j} m_j^{\dagger 2} - \text{h.c.}]$, denotes the $j^{th}$ magnon-squeezing, characterized by the squeezing strength $\Gamma_j$ and phase $\theta_j$, where $\theta_j$ represents the phase of the parametric drive generating the $j^{\text{th}}$ magnon squeezing \cite{add1}. Moreover, several experimental routes have been proposed to realize this magnon squeezing \cite{B6}, including two-tone microwave driving \cite{B9}, nonlinear magnetostrictive interactions \cite{B5}, or intrinsic magnetic anisotropy \cite{B3,D7}.

The optical cavity supports a single optical mode that couples two YIG microbridges through the mirrors attached to their surfaces. The $j^{\text{th}}$ optomechanical coupling strength is given by
\begin{equation}
	\label{aj1}
	\bar{g}_{0j}=-(-1)^{j}g_{0}\cos^{2}\left(\Psi/2\right),
\end{equation}
$\Psi$ is the angle between the incident and reflected light at the surface of the YIG-microbridge mirror \cite{D3}. The parameter $g_{m_j}$ denotes the $j^{\text{th}}$ magnomechanical coupling strength. The cavity--laser drive amplitude is $ \Upsilon=\sqrt{2\kappa_{c}\mathcal{P}_{L}/(\hbar\omega_{L})}$, where $\kappa_{c}$ is the optical cavity decay rate and $\mathcal{P}_{L}$ is the input laser power.\\

By taking into account the dissipation and the corresponding input noise for each mode, one can obtain the following quantum Langevin equations (QLEs):
\begin{eqnarray}
	\label{03}
	\label{2}
	\dot{c}&=& -i{\Delta}_{c}c-{\kappa}_{c}c-i\sum_{j=1,2}{\bar{g}_{0j}}cq_{j}+\Upsilon + \sqrt{2\kappa_{c}} c^{in},\nonumber\\
	\dot{m}_j &=& -i{\Delta}_{m_j}m_j-\kappa_{m_j}m_j-ig_{m_j}m_jq_j+\Omega_j \nonumber\\
	&& +\Gamma_{j} m_j^\dagger e^{i\theta_j}+\sqrt{2\kappa_{m_j}} m^{in}_j,\nonumber\\
	\dot{p}_j&=&-\omega_{b_j}q_j-{g_{m_j}}m_j^\dagger m_j-\bar{g}_{0j}c^\dagger c-\gamma_{b_j} p_j+\xi_j,\nonumber\\
	\dot{q}_j&=&\omega_{b_j}p_j,
\end{eqnarray}
where $\kappa_{m_j}$ defines the intrinsic decay rate of the magnon$_j$ mode and $\gamma_{b_j}$ is the damping rate of the phonon$_j$ mode. The corresponding zero-mean input noise operators $\varphi^{in}$ obey the following correlation functions \cite{D8}
\begin{equation}
	\braket{\varepsilon^{in\dagger}(t)\varepsilon^{in}(t') ; \varepsilon^{in}(t)\varepsilon^{in\dagger}(t')} = (\bar{n}_\varepsilon ; \bar{n}_\varepsilon + 1) \delta(t - t').
\end{equation} 
Moreover, $\xi_j$ denotes the zero-mean Brownian noise operator of the $j^{\text{th}}$ mechanical mode.  Under the Markovian approximation for a mechanical mode with a high quality factor $\mathcal{Q}_j = \omega_{b_j} / \gamma_{b_j} \gg 1$, it can be expressed as:
\begin{equation}
	\braket{\xi_j(t)\xi_k(t') + \xi_j(t')\xi_k(t)} = 2 \gamma_{b_j} (2\bar{n}_{b_j} + 1) \delta_{jk} \delta(t - t') , 
\end{equation}
$\bar{n}_\varepsilon = (\exp({\hbar\omega_\varepsilon}/{(k_B T)}) - 1)^{-1}$ corresponds to the average thermal occupancy of the $\varepsilon$ mode ($\varepsilon = c, m_j, b_j$) at bath temperature $T$, where $k_B$ is the Boltzmann constant \cite{D9,E0}.\\

The use of strong coherent excitation fields for the optical cavity and the magnon modes generates large steady-state amplitudes $|\alpha_s|, |m_{js}| \gg 1$. Under this condition, we can linearize the QLEs reported in Eq.~(\ref{2}) around the steady-state values $\mathcal{O}_s$. We write each operator as ${\mathcal{O}} = {\mathcal{O}}_s + \tilde{\mathcal{O}}$, with $\mathcal{O} = c, m_j, q_j, p_j$, and we neglect the second-order fluctuation terms. After linearization, the QLEs can be expressed in terms of quadrature fluctuations, which are defined as 
\begin{equation}
	\tilde{A}_\varepsilon={ (\tilde{\varepsilon} +\tilde{\varepsilon}^\dagger)/}{\sqrt{2}}\quad,\quad \tilde{B}_\varepsilon=i{(\tilde{\varepsilon}^\dagger-\tilde{\varepsilon})/}{\sqrt{2}},
\end{equation}
where $\varepsilon = c, m_j$. Hence, the dynamics takes the following matrix form:
\begin{equation}
	\label{5}
	\dot{\digamma}(t) = \chi\digamma(t) + \mathcal{N}(t),
\end{equation}
where $\digamma(t)$ is the vector of quadrature fluctuation operators, which we choose as 
\begin{equation}
	\digamma(t) = (\tilde{\sigma}_c, \tilde{\sigma}_{m_1} \oplus [\tilde{q}_1, \tilde{p}_1], \tilde{\sigma}_{m_2}\oplus [\tilde{q}_2, \tilde{p}_2])^T,
\end{equation}
and the noise vector $\mathcal{N}(t)$ is given by 
\begin{equation}
	\mathcal{N}(t) = [\sqrt{2\kappa_c}\tilde{\sigma}^{\text{in}}_c,\sqrt{2\kappa_{m_1}} \tilde{\sigma}^{\text{in}}_{m_1} \oplus [0, \xi_1], \sqrt{2\kappa_{m_2}} \tilde{\sigma}^{\text{in}}_{m_2}\oplus [0, \xi_2]]^T,
\end{equation}
with $\tilde{\sigma}_{\varepsilon} = [\tilde{A}_\varepsilon, \tilde{B}_\varepsilon]$ and $\tilde{\sigma}_{\varepsilon}^{\text{in}} = [\tilde{A}_\varepsilon^{\text{in}}, \tilde{B}_\varepsilon^{\text{in}}]$ for $\varepsilon = c, m_j$. The matrix $\chi$ is the drift matrix, which takes the following compact form:
\begin{equation}
	\label{6}
	\chi = \begin{pmatrix}
		\chi_c & \bar{\mathcal{G}} & -\bar{\mathcal{G}} \\
		\mathcal{G} & \chi_{m_1} & \mathbf{0}_{4,4} \\
		-\mathcal{G} & \mathbf{0}_{4,4} & \chi_{m_2}
	\end{pmatrix},
\end{equation}
where 
\begin{equation}
	\label{7}
	\chi_c = \begin{pmatrix}
		-\kappa_c & \tilde{\Delta}_c \\
		-\tilde{\Delta}_c & -\kappa_c
	\end{pmatrix},
\end{equation}
\begin{equation}
	\label{8}
	\chi_{m_j} = \begin{pmatrix}
		-\kappa_{m_j} +\kappa_{\theta_j} & \bar{\Delta}_{m_j} +\Delta_{\theta_j} & -G_{m_j} & 0 \\
		-\bar{\Delta}_{m_j} + \Delta_{\theta_j} & -\kappa_{m_j} - \kappa_{\theta_j} & 0 & 0 \\
		0 & 0 & 0 & \omega_{b_j} \\
		0 & G_{m_j} & -\omega_{b_j} & -\gamma_{b_j}
	\end{pmatrix},
\end{equation}
\begin{equation}
	\label{9}
	\mathcal{G}^T = \begin{pmatrix}
		0 & 0 & 0 & G_0 \\
		0 & 0 & 0 & 0
	\end{pmatrix}, \quad
	\bar{\mathcal{G}} = \begin{pmatrix}
		0 & 0 & -G_0 & 0 \\
		0 & 0 & 0 & 0
	\end{pmatrix}.
\end{equation}
The matrix $\mathbf{0}_{l,c}$ in Eq.~(\ref{6}) denotes the zero matrix of $l$ rows and $c$ columns.
Here, $\tilde{\Delta}_c = \Delta_c + \sum_{j=1,2} \bar{g}_{0j} q_{js}$, 
$\bar{\Delta}_{m_j} = \Delta_{m_j} + {g_{m_j}} q_{js}$ is the effective magnon detuning including the frequency shift due to magnomechanical interaction, and $\Delta_{\theta_j} = \Gamma_j \sin\theta_j$ is the frequency shift induced by the magnon squeezing. Typically, these frequency shifts are small, i.e., \cite{B5}
\begin{equation}
	|\bar{\Delta}_{m_j} - \Delta_{m_j}| \ll \Delta_{m_j}.
\end{equation}
Besides, $\kappa_{\theta_j} = \Gamma_j \cos\theta_j$ defines the effective magnon decay rate due to the magnon squeezing effect.
Additionally, $G_0 = i\sqrt{2} g_0 \alpha_s \cos^2(\Psi/2)$ and $G_{m_j} = i\sqrt{2} g_{m_j} m_{js}$ are the effective optomechanical and magnomechanical coupling forces, respectively, where
\begin{eqnarray}
	\label{4}
	q_{js} &=& -\frac{1}{\omega_{b_j}} \left( \bar{g}_{0j} |\alpha_s|^2 + g_{m_j} |m_{js}|^2 \right),
\end{eqnarray}
with $\alpha_s$ and $m_{js}$ are given as: 
\begin{eqnarray}
		\label{04}
		\alpha_s &=& \frac{\Upsilon}{i\tilde{\Delta}_c + \kappa_c},~~
		m_{js} = \frac{\varkappa+\Gamma_{j} e^{i\theta_j}}{|\varkappa|^2 - \Gamma_j^2} \Omega_j,
\end{eqnarray}
where $\varkappa = \kappa_{m_j} - i\bar{\Delta}_{m_j}$.

The drift matrix is derived under the condition $|\tilde{\Delta}_{c}|,|\bar{\Delta}_{m_j}|\simeq\omega_{b_j}\gg \kappa_{c},\kappa_{m_j}$ \cite{A8}, which corresponds to an optimal regime for generating entanglement in the system. Hence, the Eq. (\ref{04}) can be simplified as: 
	\begin{eqnarray}
		\label{05}
		\alpha_s = -i\frac{\Upsilon}{\tilde{\Delta}_c}, ~ m_{js} = -i\frac{\Omega_j}{\bar{\Delta}_{m_j}} \frac{1 + i\Gamma_j e^{i\theta_j}/\bar{\Delta}_{m_j}}{1 - (\Gamma_j/\bar{\Delta}_{m_j})^2}.
	\end{eqnarray}
Note that the unsqueezed case is recovered by setting $\Gamma_j = 0$~\cite{D3}.
Owing to the linearized dynamics and the Gaussian nature of the quantum noise, the steady state of the system is fully determined by its second-order moments. These are conveniently characterized by the covariance matrix {$\mathcal{V}$}, such that its elements are defined as: 
\begin{equation}
	\begin{aligned}
	{\mathcal{V}_{ij}(t)}
	=\frac{1}{2}\Big\langle
	&{\digamma_{i}(t)\digamma_{j}(t')} +{\digamma_{j}(t')\digamma_{i}(t)}
	\Big\rangle .
	\end{aligned}
\end{equation}
Then, the steady-state covariance matrix {$\mathcal{V}$} is obtained by solving the following Lyapunov equation: 
\begin{equation}
	\label{10}
	\chi\mathcal{V}+\mathcal{V}\chi^T=-\Theta,
\end{equation}
where $\Theta$ is the diffusion matrix defined by 
\begin{equation}
	\Theta_{ij}\delta(t'-t)=\frac{1}{2}\braket{n_i(t)n_j(t')+n_j(t')n_i(t)}.
\end{equation}
Additionally, it can be rewritten as 
\begin{equation}
	\Theta=\Theta_c\oplus\Theta_{m_1}\oplus\Theta_{m_2},
\end{equation}
where $\Theta_{c}=\text{diag} [{\kappa}_{c},\kappa_{c}]$ and $\Theta_{m_j}=\text{diag} [\kappa_{m_j}(2\bar{n}_{m_j}+1),\kappa_{m_j}(2\bar{n}_{m_j}+1),0,\gamma_{b_j}(2\bar{n}_{b_j}+1)]$.

\section{RESULTS AND DISCUSSION}
\label{sec:results}

To quantify quantum entanglement among the optomagnonic modes in the proposed system, we evaluate the logarithmic negativity $E_{N}$ \cite{E1,E2} and the residual contangle $\mathcal{R}_{N}$ \cite{E3,E4}, both are calculated from the covariance matrix $\mathcal{V}$ obtained from Eq.~(\ref{10}).

The bipartite entanglement between subsystems $\alpha$ and $\beta$ (where $\alpha, \beta \in \{c,m_1,m_2\}$ and $\alpha \neq \beta$) is quantified by the logarithmic negativity \cite{E1,E2}:
\begin{equation}
	E_N^{\alpha|\beta} \equiv \max[0, -\ln(2\tilde{\nu}^-)],
\end{equation}
where $\tilde{\nu}^- = \min\left|\text{eig}[i\boldsymbol{\Omega}_2\tilde{\mathcal{V}}_4]\right|$ (with the symplectic matrix $\boldsymbol{\Omega}_2 = \bigoplus_{k=1}^{2} i\sigma_y$ and the Pauli-$y$ matrix $\sigma_y$) is the smallest symplectic eigenvalue of the partially transposed covariance matrix $\tilde{\mathcal{V}}_4 = \mathcal{P}_0 \mathcal{V}_4 \mathcal{P}_0$. Here, $\mathcal{V}_4$ is the $4 \times 4$ submatrix obtained by tracing out the remaining mode from $\mathcal{V}$, and $\mathcal{P}_0 = \text{diag}(1, -1, 1, 1)$ realizes the partial transposition.\\
To characterize genuine tripartite entanglement, we use the \textit{minimum} residual contangle \cite{E5}:
\begin{equation}
	\mathcal{R}_{N}^{\min}\equiv \min\big[\mathcal{R}_{N}^{c|m_1m_2},\mathcal{R}_{N}^{m_1|cm_2},\mathcal{R}_{N}^{m_2|cm_1}\big],
\end{equation}
where $\mathcal{R}_{N}^{\alpha|\beta\delta}\equiv C_{\alpha|\beta\delta}-C_{\alpha|\beta}-C_{\alpha|\delta}$ is the residual contangle, and $C_{\alpha|\beta} \equiv [E_{N}^{\alpha|\beta}]^2$ is the contangle defined as the squared logarithmic negativity \cite{E3,E4}. This satisfies the monogamy inequality $C_{i|jk}\ge C_{i|j}+C_{i|k}$, and a nonzero $\mathcal{R}_{N}^{\min}>0$ confirms genuine tripartite entanglement.
For bipartitions where $\beta$ contains two modes, the symplectic eigenvalue is $\tilde{\nu}_{\alpha|\beta\delta} = \min\left|\text{eig}[i\boldsymbol{\Omega}_3\tilde{\mathcal{V}}_6]\right|$, with $\boldsymbol{\Omega}_3 = \bigoplus_{k=1}^{3} i\sigma_y$ and $\tilde{\mathcal{V}}_6 = \mathcal{P}\mathcal{V}\mathcal{P}$. The partial transposition matrices for the three bipartitions are:
\begin{eqnarray}
	\mathcal{P}_{c|m_1 m_2} &=& \text{diag}(1, -1, 1, 1, 1, 1), \nonumber\\
	\mathcal{P}_{m_1|c m_2} &=& \text{diag}(1, 1, 1, -1, 1, 1), \nonumber\\
	\mathcal{P}_{m_2|c m_1} &=& \text{diag}(1, 1, 1, 1, 1, -1).
\end{eqnarray}

Furthermore, to quantify the nonreciprocal of entanglement \cite{C9}, we introduce the bidirectional contrast ratio for both bipartite and tripartite correlations:
\begin{eqnarray}
	\label{contrast}
	{C}_{{E}}&=& \frac{|{E}_{{N}}(\theta)-{E}_{{N}}(\theta+\pi)|}{{E}_{{N}}(\theta)+{E}_{{N}}(\theta+\pi)},\\ \nonumber
	{C}_{\mathcal R}&=& \frac{|\mathcal{R}_{{N}}(\theta)-\mathcal{R}_{{N}}(\theta+\pi)|}{\mathcal{R}_{{N}}(\theta)+\mathcal{R}_{{N}}(\theta+\pi)}.
\end{eqnarray}
Here, $\theta$ and $\theta+\pi$ correspond to opposite squeezing phases, which induce opposite signs for both the frequency shift ($\Delta_{\theta_j} = \Gamma_j \sin\theta_j$) and the damping shift ($\kappa_{\theta_j} = \Gamma_j \cos\theta_j$) of the magnon mode. Analogous to the contrast ratios defined in Refs.~\cite{C9,D1}, here $\theta \in (0,\pi)$ corresponds to $\Delta_{\theta_j} > 0$, while $\theta \in (\pi, 2\pi)$ corresponds to $\Delta_{\theta_j} < 0$, cf. Fig.~\ref{fig:1}(c).
The value ${C}_{{E}}$ (${C}_{\mathcal R}$) $=1$ represents ideal nonreciprocity, while ${C}_{{E}}$ (${C}_{\mathcal R}$) $=0$ indicates complete absence of nonreciprocity for bipartite (tripartite) entanglement. When both entanglement values in the denominator vanish, the corresponding contrast ratio is mathematically undefined. In the density plots below, these zero-denominator points are left blank or white and are not interpreted as ideal nonreciprocal entanglement.

Our ring-cavity configuration combines two ferrimagnetic YIG microbridges with a single optical mode, creating a system where magnons, phonons, and photons interact through radiation pressure and magnetostriction. The mechanical modes, operating at megahertz frequencies, remain highly thermally populated even in a dilution refrigerator. Therefore, they must be cooled to a temperature close to the ground state to allow the preparation of quantum states of the system. To achieve this, we drive the optical cavity with a red-detuned laser field at $\omega_c - \omega_b$, which is resonant with the optomechanical anti-Stokes sideband. In the absence of magnon squeezing, one can prepare optomagnonic entangled states following the protocol reported in Ref.~\cite{C3}.

Here, we introduce a squeezing process to achieve switchable control through two experimental scenarios. In scenario (I), the first YIG microbridge is driven with a red-detuned microwave field at frequency $\omega_{m_1} - \omega_b$ to achieve magnon squeezing. Meanwhile, the second magnon remains unsqueezed and is driven with a blue-detuned microwave field at $\omega_{m_2} + \omega_b$. This creates a striking asymmetry: the squeezed magnon exhibits phase-dependent changes in frequency shift and damping, while the unsqueezed magnon generates entanglement through parametric down-conversion. In scenario (II), we reverse the roles, such that the second magnon is squeezed and the first one is not. Note that this swap (scenario I vs scenario II) is independent of the squeezing phase reversal $\theta \to \theta+\pi$ that produces the entanglement nonreciprocity. It demonstrates an additional switchability between the entanglement channels $E_N^{c|m_2}$ and $E_N^{c|m_1}$ by selecting which magnon is squeezed.

The squeezing process shifts the effective magnon frequency by $\Delta_{\theta_j} = \Gamma_j \sin\theta_j$ and modifies its decay rate by $\kappa_{\theta_j} = \Gamma_j \cos\theta_j$. When the squeezing phase is reversed from $\theta$ to $\theta + \pi$, both quantities reverse their signs, analogous to reversing the Kerr coefficient sign \cite{C9} or flipping the Barnett-induced frequency shift via magnetic field direction \cite{D1}. However, conventional approaches require either mechanical rotation, magnetic field engineering, or intrinsic nonlinearities. Our method only requires adjusting the squeezing phase.
The considered simulation parameters are consistent with the values reported in Refs. \cite{D3,A1,D04} (see Table. \ref{tab:1}).
\begin{table}[h]
	\captionsetup{justification=RaggedRight, singlelinecheck=false}
	\caption{The parameters used in our simulation (taken from recent reported experiments).}
	\centering
	\begin{tabular}{lc}
		\hline\hline
		Parameter & Value \\
		\hline
		\multicolumn{2}{l}{\textit{\underline{Optical cavity:}}} \\
		~~~~Optical wavelength & ~~$\lambda_L = 1064$ nm \\
		~~~~Cavity decay rate & ~~$\kappa_c/2\pi = 3$ MHz \\	
		~~~~Angle & ~~$\Psi = \pi/3$ \\
		~~~~Effective OM coupling &~~ $G_0/2\pi = 8$ MHz \\ [0.3em]
		\multicolumn{2}{l}{\textit{\underline{YIG microbridge of volume $5 \times 2 \times 1~\mu\text{m}^3$:}}} \\
		~~~~Magnon frequency &~~ $\omega_m/2\pi = 10$ GHz \\
		~~~~Mechanical frequency &~~ $\omega_b/2\pi = 40$ MHz \\
		~~~~Magnon$_1$ decay rate &~~ $\kappa_{m_1}/2\pi = 1$ MHz \\
		~~~~Magnon$_2$ decay rate & ~~$\kappa_{m_2} = \kappa_{m_1}/2$ \cite{ZI1} \\
		~~~~Mechanical damping & ~~$\gamma_b/2\pi = 10^2$ Hz \\
		~~~~Bath temperature &~~ $T = 10$ mK \\
		\hline\hline
	\end{tabular}
	\label{tab:1}
\end{table}

The effective optomechanical coupling $G_0/2\pi = 8$ MHz is achieved with laser power $\mathcal{P}_L = 17.80$ mW for $g_0/2\pi = 1$ kHz at detuning $\tilde{\Delta}_c = \omega_b$ and $\Psi = \pi/3$. The angle $\Psi$ controls the optomechanical coupling strength via $\bar{g}_{0j} \propto \cos^2(\Psi/2)$ (see Eq.~\eqref{aj1}). For $\Psi = \pi/3$, $\cos^2(\Psi/2) = 3/4$, which is a typical value in ring-cavity setups \cite{D3}.
In the regime where $|\bar{\Delta}_{m_j}| \simeq \omega_{b_j} \gg \kappa_{m_j}$ and $\Gamma_j = k\omega_b$ with $0 \le k \le 1$, the effective magnomechanical coupling becomes $G_{m_j} = \sqrt{2} g_m \Omega_j/\omega_b \, [1 + ike^{i\theta_j}]/[1-k^2]$ (see Eq.~\ref{05}).
Indeed, for magnomechanical parameters $\Gamma_j/\omega_b = 0.4$ ($k = 0.4$), and $g_m/2\pi \simeq 20$ Hz \cite{E6}, achieving $G_{m_j}/2\pi \sim 5$ MHz requires driving magnetic fields $H_{dj} = 3.07 \times 10^{-3}$ T at $\theta_j = \pi/2$ (corresponding to microwave power $\mathcal{P}_{m_j} = 11.31$ mW) and $H_{dj} = 1.32 \times 10^{-3}$ T at $\theta_j = 3\pi/2$ (corresponding to $\mathcal{P}_{m_j} = 2.08$ mW) \cite{E7}.
In the unsqueezed case, achieving $G_{m_j}/2\pi \sim 1$ MHz requires $H_{dj} = 3.96 \times 10^{-4}$ T (corresponding to $\mathcal{P}_{m_j} = 0.19$ mW) at detuning  $\bar{\Delta}_m = -0.9 \omega_b$. These drive powers lie well within the milliwatt-to-hundred-milliwatt range accessible in current cavity magnomechanics experiments using loop antennas. 
All key parameters used in this work are grounded in current experimental platforms, the OM coupling $g_0/2\pi =1$~kHz for the micron-sized mirror \cite{D5} and the MM coupling $g_m/2\pi \simeq 20$~Hz for the YIG microbridge geometry \cite{E6,D04} are both realized within the ring-cavity optomagnomechanical architecture \cite{D3}.
Note that radiation-pressure and magnetostrictive couplings depend on drive powers \cite{E7}, explicitly
\begin{eqnarray}
	\mathcal{P}_L &=& \frac{\hbar\omega_L}{\kappa_c} 
	\bigg[\frac{\tilde{\Delta}_c}{2g_0\cos^2(\Psi/2)}\bigg]^2 G_0^2,\nonumber\\
	\mathcal{P}_{m_j} &=& \frac{LWc}{\rho\,\mu_0\,V} 
	\bigg[\frac{2\bar{\Delta}_{m_j}}{\sqrt{5}g_{m_j}\,\gamma}\bigg]^2 
	\frac{G_{m_j}^2}{|\mathcal{L}_s|^2},
\end{eqnarray}
where $\mathcal{L}_s = [1 + i\Gamma_j e^{i\theta_j}/\bar{\Delta}_{m_j}]/[1 - (\Gamma_j/\bar{\Delta}_{m_j})^2]$. This separate tunability of $G_0$ via the laser power $\mathcal{P}_L$ and $G_{m_j}$ via the microwave drive $\mathcal{P}_{m_j}$ is a distinctive feature of the ring-cavity OMM architecture, unavailable in standard cavity magnomechanics where the absence of an optical mode prevents independent control of the OM and MM coupling channels. All results are obtained in stable regimes verified using the so-called Routh--Hurwitz criterion (see Appendix~B) \cite{E8}.

\begin{figure}[h]
	\begin{center}
		\subfigure{\includegraphics[scale=0.22]{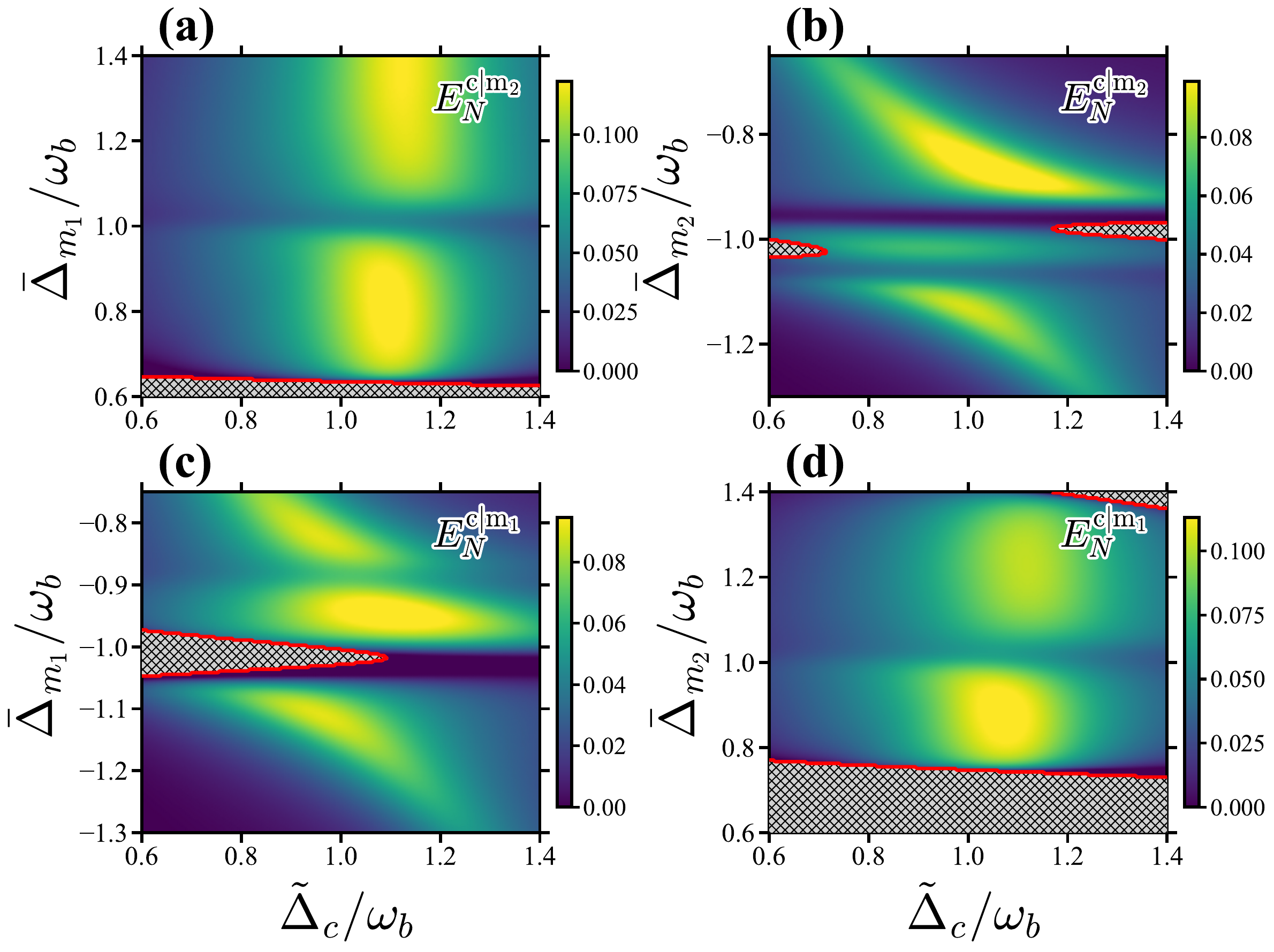}}
	\end{center}
	\vspace{-1.5em} 
	\captionsetup{justification=RaggedRight, singlelinecheck=false}
	\caption{{\small Density plots of cavity-magnon entanglement (a-b) $E_N^{c|m_2}$ for scenario I and (c-d) $E_N^{c|m_1}$ for scenario II versus (a,c) $(\tilde{\Delta}_c, \Delta_{m_1})$, (b,d) $(\tilde{\Delta}_c, \Delta_{m_2})$, where $\tilde{\Delta}_c$ is the effective cavity detuning and $\Delta_{m_j}$ is the intrinsic detuning of magnon $j$. The red boundary indicates stability.  For scenario I, we set: $G_{m_1}/2\pi = 5$ MHz, $G_{m_2}/2\pi = 1$ MHz. While, in scenario II, we set $G_{m_1}/2\pi = 1$ MHz, $G_{m_2}/2\pi = 5$ MHz. See Table~.\ref{tab:1} for other parameters.}}
	\label{fig:2}
\end{figure}

To identify optimal operating regimes, we plot the amount of entanglement in Fig.~\ref{fig:2}. Panels (a,b) show scenario I, where magnon$_1$ is squeezed. Panel (a) exhibits a broad plateau centered at $\tilde{\Delta}_c \simeq\omega_b$, arising from the optomechanical anti-Stokes resonance that cools the mechanical mode and for $\bar{\Delta}_{m_1}\simeq 1.1\omega_b$ sustains cavity-magnon$_2$ correlations over a wide detuning range. Panel (b) displays a much narrower resonance near to $\tilde{\Delta}_c\simeq 1.1\omega_b$, $\Delta_{m_2} \simeq -0.9\omega_b$, originating from the simultaneous resonance of anti-Stokes optomechanical and Stokes magnomechanical parametric down-conversion, which restricts entanglement to a sharp double-resonance region.
Panels (c,d) examine scenario II with reversed detunings $\Delta_{m_2} \simeq \omega_b$, demonstrating the role of the squeezed magnon selection. All entanglement regions lie within stability boundaries (red contours), ensuring experimental accessibility and validity of the proposed model.

\begin{figure}[t]	
	\begin{center}
		\subfigure{\includegraphics[scale=0.18]{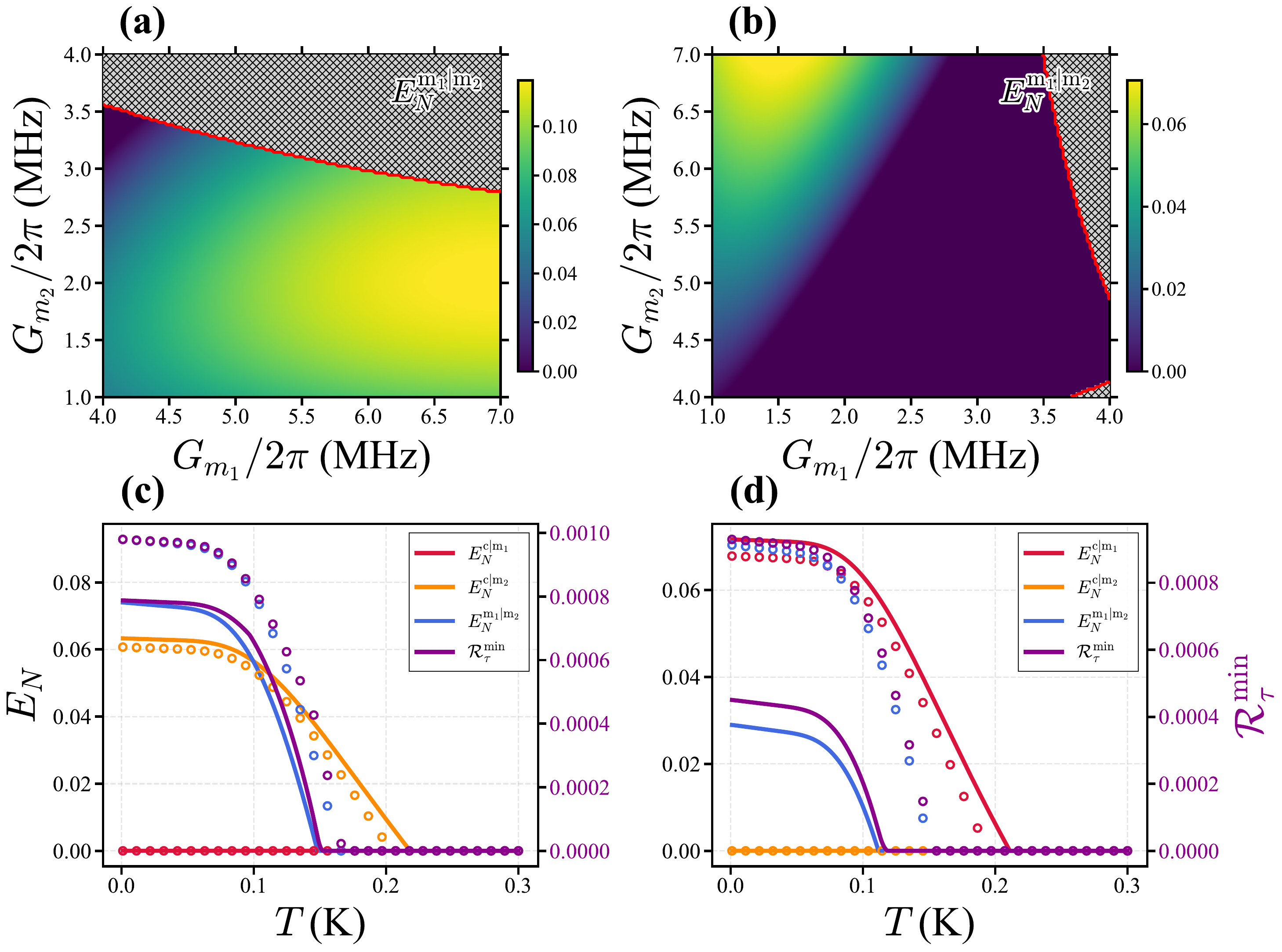}}
	\end{center}
	\vspace{-2em} 
	\captionsetup{justification=RaggedRight, singlelinecheck=false}
	\caption{{\small Density plots of magnon-magnon entanglement $E_N^{m_1|m_2}$ versus $(G_{m_1}, G_{m_2})$ for (a) scenario I and (b) scenario II. Temperature dependence for (c) scenario I and (d) scenario II showing cavity-magnon entanglement $E_N^{c|m_1}$ (red), $E_N^{c|m_2}$ (orange), magnon-magnon entanglement $E_N^{m_1|m_2}$ (blue), and tripartite entanglement {$\mathcal{R}_{N}^{\min}$} (purple) versus bath temperature $T$. Moreover, we set $\theta = \pi/2$ and $\theta = 3\pi/2$ for solid and circles lines, respectively. Other parameters are the same as in Fig.~\ref{fig:2}.}}
	\label{fig:3}
\end{figure}

\begin{figure}[t]
	\begin{center}
		\subfigure{\includegraphics[scale=0.23]{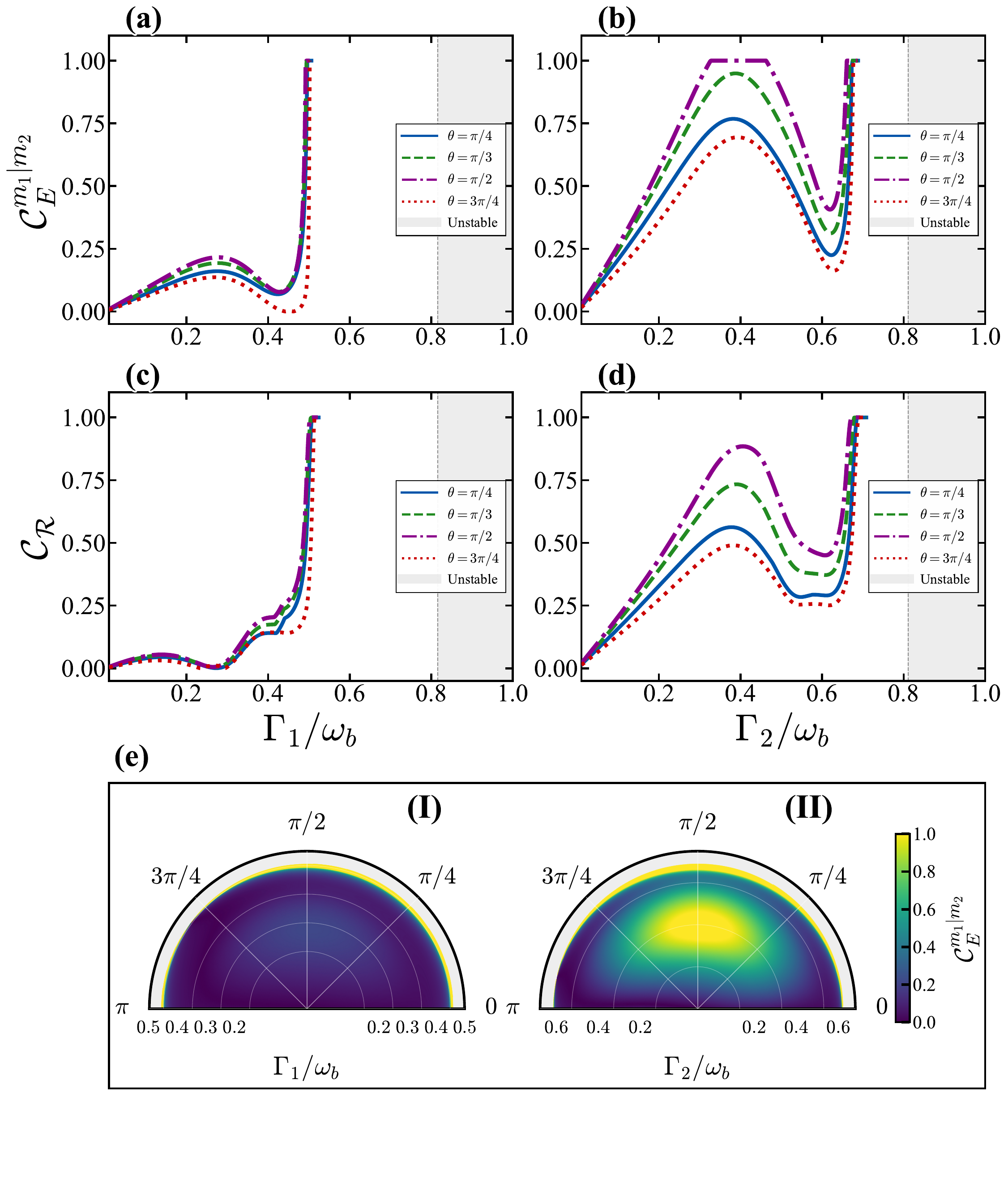}}
	\end{center}
	\vspace{-4.em} 
	\captionsetup{justification=RaggedRight, singlelinecheck=false}
	\caption{{\small Nonreciprocal contrasts $C_E^{m_1|m_2}$ and $C_{\mathcal R}^{\min}$ for (a,c) scenario I and (b,d) scenario II versus $\Gamma_{1,2}/\omega_b$. Nonreciprocity is revealed by comparing $\theta$ and $\theta + \pi$, with $\theta = \pi/4$, $\pi/3$, $\pi/2$, and $3\pi/4$. (e) Density plots of $C_E^{m_1|m_2}$ as a function of $(\theta, \Gamma_j/\omega_b)$ for scenario I and scenario II. Other parameters are the same as in Fig.~\ref{fig:2}.}}
	\label{fig:4}
\end{figure}

To ensure the system's stability, we set $G_0 > G_{m_j}$, allowing the optomechanical anti-Stokes process (phonon absorption) to dominate magnomechanical Stokes emission. Additionally, the unsqueezed magnon requires stronger coupling than the squeezed one to preserve stability, as illustrated in Figs.~(\ref{fig:3}a,\ref{fig:3}b), which display magnon-magnon entanglement $E_N^{m_1|m_2}$ versus coupling strengths, revealing broad stable regions with optimal values at moderate asymmetry $G_{m_j}/ 2\pi \simeq 4$--$7$ MHz.
Figs.~(\ref{fig:3}c, \ref{fig:3}d) examine the temperature dependence at opposite phases. In panel (c) (scenario I), at $\theta_1 = \pi/2$, cavity-magnon entanglement $E_N^{c|m_2}$ survives up to $T \simeq 218$ mK while $E_N^{c|m_1}$ vanishes, demonstrating phase-dependent suppression. 
This asymmetric thermal tolerance reflects the phase-controlled modification of both the effective magnon frequency shift $\Delta_\theta = \Gamma_j \sin\theta$ and the dissipation $\kappa_\theta = \Gamma_j \cos\theta$, which selectively protects one cavity-magnon channel while degrading the other. Magnon-magnon entanglement $E_N^{m_1|m_2}$ persists up to $T \simeq 149$ mK, and tripartite entanglement $\mathcal{R}_{N}^{\min}$ up to $T \simeq 152$ mK. Flipping to $\theta_1 = 3\pi/2$ reverses the pattern, with $E_N^{c|m_2}$ surviving up to $T \simeq 207$ mK, $E_N^{m_1|m_2}$ up to $T \simeq 166$ mK, and $\mathcal{R}_{N}^{\min}$ up to $T \simeq 176$ mK. 
The enhanced thermal robustness of cavity-magnon entanglement compared to magnon-magnon and tripartite entanglement reflects the dominant role of phonon thermal occupation $\bar{n}_{b_j} \simeq k_B T/(\hbar \omega_{b_j})$ at MHz frequencies, which primarily degrades phonon-mediated correlations. Panel (d) mirrors this behavior in scenario II. At $\theta_2 = \pi/2$, $E_N^{c|m_1}$ survives up to $T \simeq 211$ mK, while $E_N^{c|m_2}$ vanishes. Magnon-magnon entanglement persists up to $T \simeq 112$ mK, and $\mathcal{R}_{N}^{\min}$ up to $T \simeq 119$ mK. At $\theta_2 = 3\pi/2$, these values shift to $T \simeq 197$ mK, $154$ mK and $155$ mK, respectively. Magnon squeezing does not eliminate the need for cryogenic operation. The mechanical modes are in the MHz range, so their thermal occupation remains large even at sub-Kelvin temperatures; cryogenic precooling and sideband cooling are therefore still required for phonon-mediated correlations. The advantage of the squeezing-induced control is that several calculated correlations remain nonzero up to temperatures of order $0.1$--$0.2$~K, with the most robust cavity--magnon case reaching about $218$~mK. The temperature intervals where entanglement appears in one phase and disappears in the other define ideal zones of nonreciprocal entanglement, confirming phase-controlled switching within dilution refrigerator reach.

\begin{figure}[t]	
	\begin{center}
		\subfigure{\includegraphics[scale=0.22]{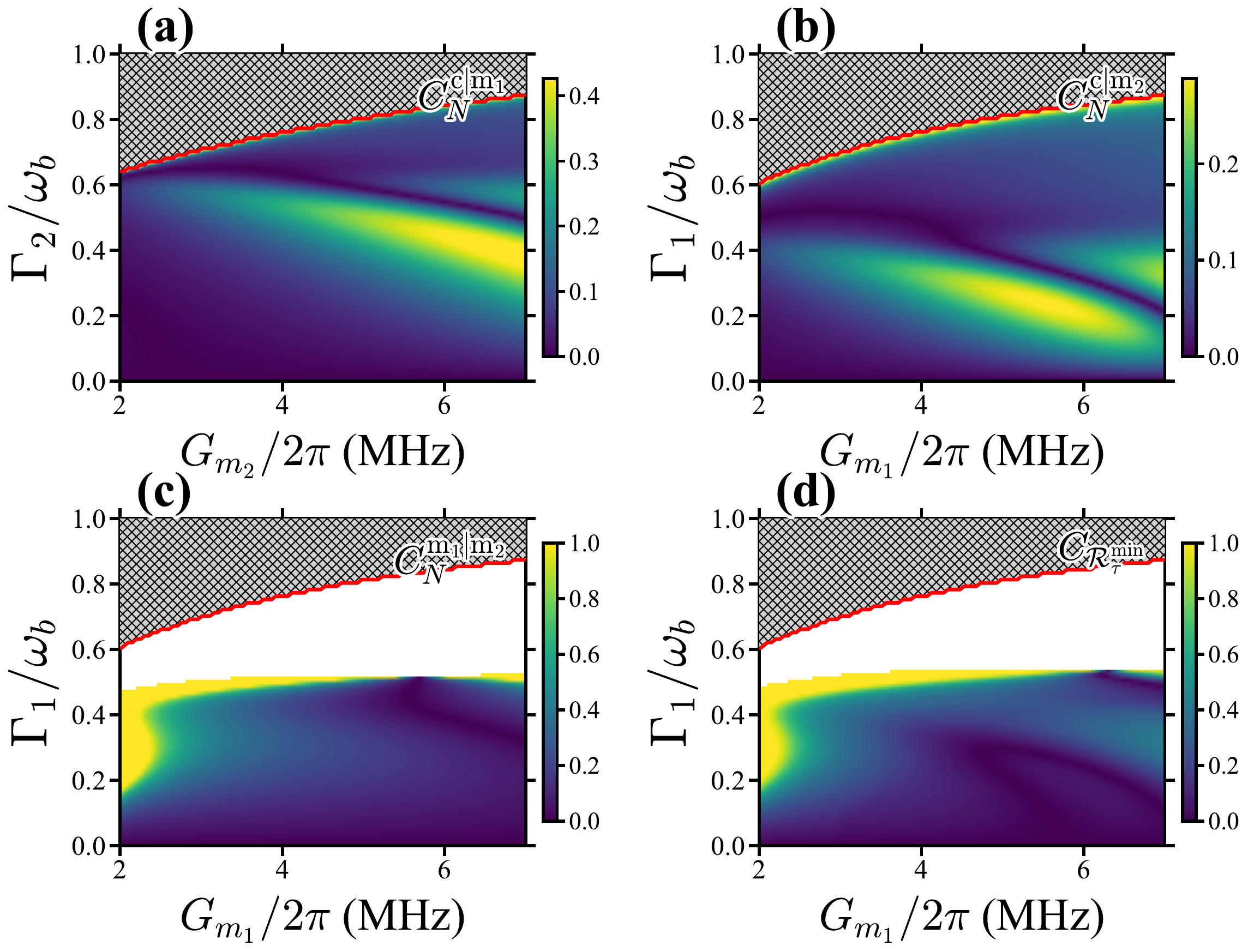}}
	\end{center}
	\vspace{-2em} 
	\captionsetup{justification=RaggedRight, singlelinecheck=false}
	\caption{{\small Nonreciprocal contrast versus coupling and squeezing. Bidirectional contrast ratios (a) $C_E^{c|m_1}$ (scenario II), (b) $C_E^{c|m_2}$ (scenario I), (c) $C_E^{m_1|m_2}$ (scenario I), and (d) $C_{\mathcal R}^{\min}$ (scenario I) versus magnomechanical coupling $G_{m_j}$ and squeezing strength $\Gamma_j$, where $j = 1, 2$ denotes which magnon is squeezed. The red boundary indicates stability. Other parameters are the same as in Fig.~\ref{fig:2}.}}
	\label{fig:5}
\end{figure}

Fig.~\ref{fig:4} extends this analysis to multiple squeezing phases $\theta = \pi/4$, $\pi/3$, $\pi/2$, and $3\pi/4$. In scenario~I, panels (a,c), ideal nonreciprocity ($C_E^{m_1|m_2},$ {$C_{\mathcal R}^{\min}$} $\simeq 1$) is achieved for all phases. At $\theta = \pi/3$, both shifts $|\Delta_{\theta}| = \sqrt{3}\Gamma_j/2$ and $|\kappa_{\theta}| = \Gamma_j/2$ contribute simultaneously, highlighting the advantage of dual-parameter control over single-frequency-shift methods. The scans suggest that broad high-contrast regions occur when the frequency-shift and quadrature-damping contributions have comparable magnitudes, $|\Delta_\theta|\sim |\kappa_\theta|$, i.e., $\tan\theta \approx 1$, as occurs near $\theta=\pi/4$ and $3\pi/4$. This should be understood as a physical guideline rather than a closed-form optimality condition, since the exact stability and entanglement windows are determined by the full drift matrix.
In scenario~II, panels (b,d), ideal nonreciprocity is reached at $\Gamma_2 \lesssim 0.5\,\omega_b$. At $\theta = \pi/2$, the dissipation shift vanishes ($\kappa_{\theta} = 0$) and the squeezing Hamiltonian reduces to $H_{\text{sq}} = \pm (\Gamma_2/2)(m_j^{\dagger 2} + m_j^2)$, where only the frequency shift drives nonreciprocity. Phases with $\theta \neq \pi/2$ activate both channels, extending ideal nonreciprocity over broader stable regions (see Appendix~B). Remarkably, squeezing the second magnon (scenario~II) yields the largest stability region and the most extensive zone of ideal nonreciprocity, making it the optimal configuration for robust switchable entanglement in this platform. Panel~(e) extends this analysis to the full $(\theta, \Gamma_j)$ space, showing that scenario~II maintains high contrast ($C_E \simeq 1$) over a finite range $\theta \in [\pi/4, 3\pi/4]$ at $\Gamma_2/\omega_b \simeq 0.3$--$0.5$. These finite high-contrast regions indicate tolerance to moderate phase deviations, coupling variations, and thermal fluctuations within the tested parameter ranges. A complete device-specific error budget would require specifying the concrete implementation of the squeezing pump and is beyond the scope of the present theoretical model.

Fig.~\ref{fig:5} quantifies nonreciprocal strength via contrast ratios. Panels (a,b) show cavity-magnon contrast $C_E^{c|m_1}$ in scenario II and $C_E^{c|m_2}$ in scenario I versus coupling and squeezing. Strong entanglement nonreciprocity emerges at $G_{m_j}/2\pi \simeq 5$--$6$ MHz for $\Gamma_2/\omega_b \simeq 0.3$--$0.6$ in scenario II and $\Gamma_1/\omega_b \simeq 0.2$--$0.4$ in scenario I. These coupling strengths can be achieved with microwave powers $\mathcal{P}_{m_j} = 11.31$--$16.29$~mW at $\theta = \pi/2$ and $\mathcal{P}_{m_j} = 2.08$--$2.99$ mW at $\theta = 3\pi/2$, for $\Gamma_j/\omega_b = 0.4$. Panel (c) displays magnon-magnon contrast $C_E^{m_1|m_2}$, reaching ideal nonreciprocity in a narrower band around $\Gamma_1/\omega_b \simeq 0.4$. Panel (d) shows tripartite contrast $C_{\mathcal R}^{\min}$, lower in magnitude due to monogamy but following similar trends as panel (c). In panels (c,d), the white regions within the stable zone indicate parameter regimes where entanglement disappears at both $\theta$ and $\theta + \pi$. 
At these points, the denominators in Eq.~\eqref{contrast} vanish; accordingly, the regions are left white and are not counted as ideal nonreciprocity. Moderate coupling asymmetry around $G_{m_1}/2\pi \simeq 3$--$7$ MHz combined with squeezing $\Gamma_1/\omega_b \simeq 0.2$--$0.6$ yields an ideal switchable entanglement, corresponding to microwave powers $\mathcal{P}_{m_1} = 4.07$--$22.18$ mW at $\theta = \pi/2$ and $\mathcal{P}_{m_1} = 0.75$--$4.07$ mW at $\theta = 3\pi/2$, for $\Gamma_j/\omega_b = 0.4$. 
Compared with frequency-shift-based Kerr and Barnett mechanisms \cite{C9,D1} and with chiral-coupling schemes based on coupling asymmetry \cite{D0}, the present scheme uses the squeezing phase to control both $\Delta_\theta$ and $\kappa_\theta$ while preserving the same device geometry. Table~\ref{tab:2} summarizes these distinctions and reports whether the compared schemes require rotation, chiral coupling, or other platform-specific control mechanisms.
\begin{widetext}
	\vspace{-1.5em}
	\begin{center}
		\begin{table}[H]
			\caption{{\small Comparison with representative mechanisms for nonreciprocal or directional entanglement. The table distinguishes the physical control mechanism, the controlled quantity, the need for rotation or a platform-specific coupling asymmetry, and the reported nonreciprocity measure. Here, ``switchable channel'' denotes whether the dominant entanglement channel can be changed at fixed device geometry by applying the squeezing pump to a different magnon mode.}}
			\label{tab:2}
			{
				\renewcommand{\arraystretch}{1.25}
				\setlength{\tabcolsep}{4pt}
				\centering
				\begin{tabular}{llllll}
					\hline\hline
					\parbox[t]{1.8cm}{\raggedright Mechanism / work} &
					\parbox[t]{2.2cm}{\raggedright Platform} &
					\parbox[t]{3.cm}{\raggedright Directionality control} &
					\parbox[t]{3.7cm}{\raggedright Controlled quantity} &
					\parbox[t]{3.8cm}{\raggedright Reported measure / requirement} &
					\parbox[t]{1.7cm}{\raggedright Switchable channel} \\
					\hline
					\parbox[t]{1.8cm}{\raggedright Rotating OM \cite{C4,C44}} &
					\parbox[t]{2.2cm}{\raggedright Optical or OM resonator} &
					\parbox[t]{3.cm}{\raggedright Mechanical rotation or counter-propagating drive direction} &
					\parbox[t]{3.7cm}{\raggedright Rotation-induced frequency shift} &
					\parbox[t]{3.8cm}{\raggedright Directional entanglement enhancement; mechanical rotation required} &
					\parbox[t]{1.7cm}{\raggedright No} \\[1em]
					\parbox[t]{1.8cm}{\raggedright Kerr-based cavity MM \cite{C9,C8}} &
					\parbox[t]{2.2cm}{\raggedright Cavity MM} &
					\parbox[t]{3.cm}{\raggedright Kerr-induced magnon response, sometimes assisted by spinning} &
					\parbox[t]{3.7cm}{\raggedright Mainly frequency shift and two-magnon effect} &
					\parbox[t]{3.8cm}{\raggedright Nonreciprocal contrast reported in the corresponding platform; requires Kerr nonlinearity or spinning control} &
					\parbox[t]{1.7cm}{\raggedright No} \\[1em]
					\parbox[t]{1.8cm}{\raggedright Barnett-effect schemes \cite{D1,D11}} &
					\parbox[t]{2.2cm}{\raggedright Cavity MM} &
					\parbox[t]{3.cm}{\raggedright Barnett-effect-induced reversal controlled by magnetic-field or rotation direction} &
					\parbox[t]{3.7cm}{\raggedright Frequency shift} &
					\parbox[t]{3.8cm}{\raggedright Bipartite and tripartite nonreciprocity; requires Barnett/rotation-related control} &
					\parbox[t]{1.7cm}{\raggedright No} \\[1em]
					\parbox[t]{1.8cm}{\raggedright Chiral-coupling scheme \cite{D0}} &
					\parbox[t]{2.2cm}{\raggedright Cavity MM with counter-propagating modes} &
					\parbox[t]{3.cm}{\raggedright Driving different circulating modes with chiral cavity--magnon coupling} &
					\parbox[t]{3.7cm}{\raggedright Coupling asymmetry} &
					\parbox[t]{3.8cm}{\raggedright Nonreciprocal bipartite and tripartite entanglement; requires a chiral-coupling platform} &
					\parbox[t]{1.7cm}{\raggedright Drive-direction only} \\[1em]
					\parbox[t]{1.8cm}{\raggedright Magnon squeezing in MM \cite{C6}} &
					\parbox[t]{2.2cm}{\raggedright Standard cavity MM} &
					\parbox[t]{3.cm}{\raggedright Squeezing-phase reversal of a single magnon} &
					\parbox[t]{3.7cm}{\raggedright Frequency contribution and damping contribution, ($\pm\Delta_\theta;\pm\kappa_\theta$)} &
					\parbox[t]{3.8cm}{\raggedright High contrast reported for that single-magnon platform} &
					\parbox[t]{1.7cm}{\raggedright No} \\[1em]
					\parbox[t]{1.8cm}{\raggedright Present work} &
					\parbox[t]{2.2cm}{\raggedright Ring-cavity OMM with two YIG microbridges} &
					\parbox[t]{3.cm}{\raggedright Squeezing-phase reversal and choice of the squeezed magnon} &
					\parbox[t]{3.7cm}{\raggedright Frequency contribution and quadrature-damping contribution, ($\pm\Delta_\theta;\pm\kappa_\theta$)} &
					\parbox[t]{3.8cm}{\raggedright $C_E\simeq1$ and $C_{\mathcal R}\simeq1$ in selected Routh--Hurwitz stable regions} &
					\parbox[t]{1.7cm}{\raggedright Yes, by squeezing either $m_1$ or $m_2$} \\
					\hline\hline
			\end{tabular}}
		\end{table}
	\end{center}
	\vspace{-2.7em}
\end{widetext}

\begin{figure}[h]	
	\begin{flushleft}
		\subfigure{\includegraphics[scale=0.26]{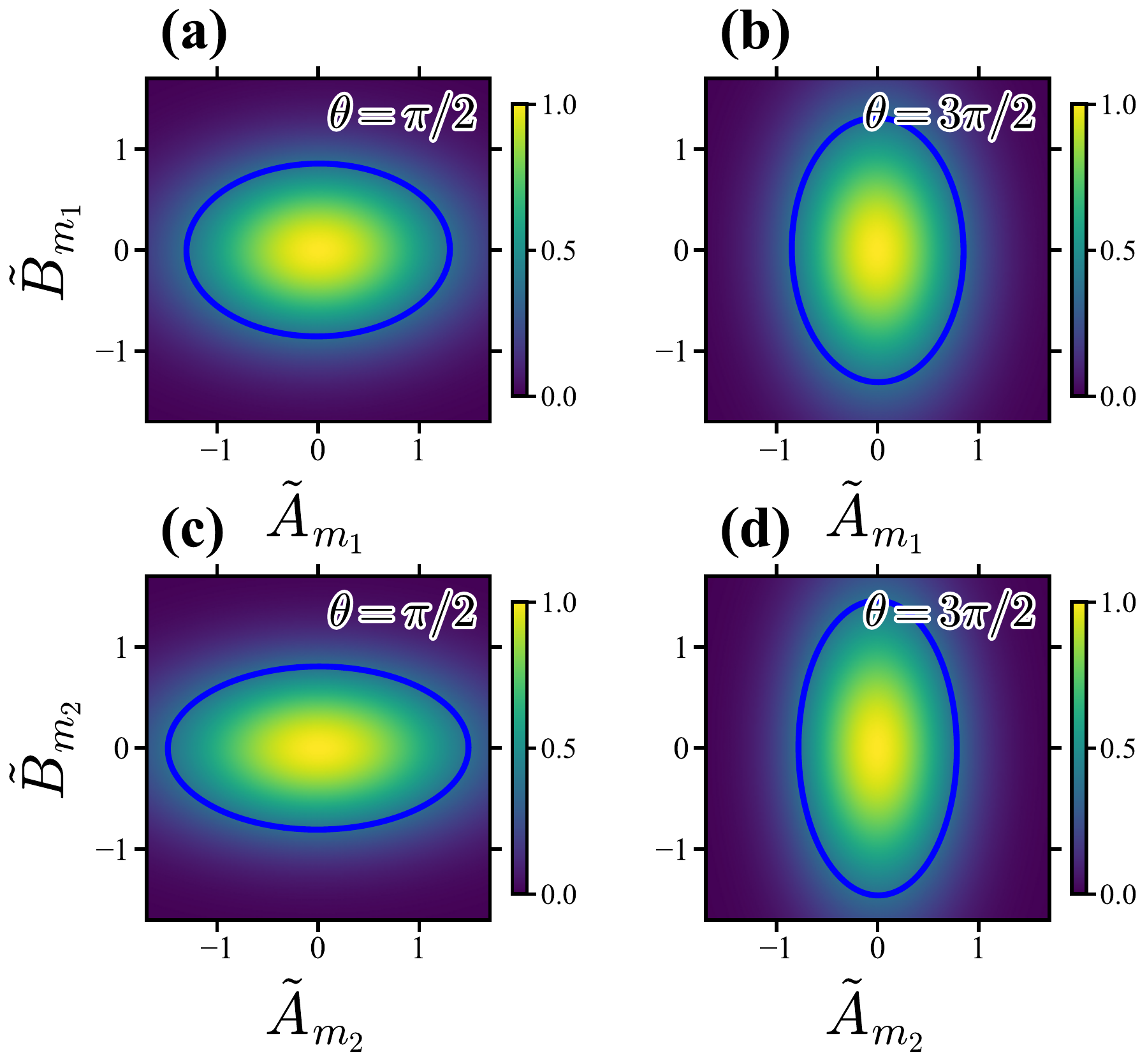}}
	\end{flushleft}
	\vspace{-1em} 
	\captionsetup{justification=RaggedRight, singlelinecheck=false}
	\caption{Reconstructed Wigner functions for (a,b) scenario I (first magnon squeezed) and (c,d) scenario II (second magnon squeezed) at $\theta = \pi/2$ (left) and $\theta = 3\pi/2$ (right). Solid blue contours indicate the $1/e$ of the maximum value of {$\mathcal{W}(\sigma)$}. Moreover, we set for scenario I: {$\bar{\Delta}_{m_1}/\omega_b = 1.1$}, {$\bar{\Delta}_{m_2}/\omega_b = -0.9$}, $\Gamma_1/\omega_b = 0.4$. While, for scenario II: {$\bar{\Delta}_{m_1}/\omega_b = -0.9$}, {$\bar{\Delta}_{m_2}/\omega_b = 1$}, $\Gamma_2/\omega_b = 0.6$.}
	\label{fig:6}
\end{figure}

To verify the magnon squeezing mechanism underlying our nonreciprocal entanglement protocol, we reconstruct the Wigner function for the magnon modes. For a Gaussian state, it takes the following form \cite{C6,C3}
\begin{equation} 
	\mathcal{W}(\sigma) = \frac{1}{\pi^2 \sqrt{\det \tilde{\mathcal{V}}}} \exp\left(-\frac{1}{2}\sigma \tilde{\mathcal{V}}^{-1} \sigma^{\dagger}\right), 
\end{equation} 
where $\sigma = (\tilde{A}_{m_j}, \tilde{B}_{m_j})$ is the magnon quadrature vector and $\tilde{\mathcal{V}}$ is the corresponding $2 \times 2$ covariance submatrix extracted from the full system covariance matrix {$\mathcal{V}$}. Fig.~\ref{fig:6} displays reconstructed Wigner functions at opposite squeezing phases. In panels (a,c) where $\theta = \pi/2$, the distributions exhibit pronounced elliptical compression along one quadrature, a clear signature of strong squeezing. Reversing to $\theta = 3\pi/2$ in panels (b,d) yields similar squeezing strength but with perpendicular orientation. This $\pi$ rotation of the squeezing axis directly visualizes how flipping $\theta$ reverses $\Delta_{\theta_j} = \Gamma_j \sin\theta_j$, confirming genuine magnon squeezing as the physical resource underlying the nonreciprocal entanglement demonstrated in Figs.~\ref{fig:2}--\ref{fig:5}.

{Finally, the above results are valid when the low-excitation assumption $\langle m_j^\dagger m_j \rangle \ll 2N_0 s = 5N_0$ is satisfied, with $s = 5/2$ for Fe$^{3+}$ in YIG. For our YIG microbridge ($V = 5 \times 2 \times 1~\mu$m$^3$), the number of spins $N_0 = 4.22 \times 10^{10}$, and $G_{m_j}/2\pi = 5$~MHz with $\Gamma_j/\omega_b = 0.4$ corresponds to $|\langle m_j \rangle| \simeq 1.77 \times 10^5$ for $\Omega_j \simeq 6.22 \times 10^{13}$~Hz at $\theta_j = \pi/2$, and for $\Omega_j \simeq 2.67 \times 10^{13}$~Hz at $\theta_j = 3\pi/2$, which leads to $\langle m_j^\dagger m_j \rangle \simeq 3.13 \times 10^{10} \ll 5N_0 = 2.1 \times 10^{11}$, and the assumption is thus well satisfied for both squeezing phases. More precisely, in the linearized Gaussian treatment the total magnon population includes both the coherent steady-state amplitude and the fluctuation contribution extracted from the covariance matrix,
\begin{equation}
\langle m_j^\dagger m_j\rangle = |m_{js}|^2 + \frac{1}{2}\left({\langle \tilde{A}_{m_j}^2\rangle + \langle \tilde{B}_{m_j}^2\rangle} -1\right).
\end{equation}
For the plotted working points, this fluctuation correction remains negligible compared with the spin-number bound $5N_0$, so the low-excitation approximation is not affected.}

\section{CONCLUSION}
\label{sec:conclusion}

In this work, we have theoretically studied phase-switchable nonreciprocal entanglement in an optomagnomechanical ring cavity using phase-controlled magnon squeezing. Reversing the squeezing phase by $\pi$ changes the signs of both the effective frequency contribution {$\Delta_{\theta_j} =\Gamma_j\sin\theta_j$} and the quadrature-damping contribution {$\kappa_{\theta_j}=\Gamma_j\cos\theta_j$}, thereby producing two dynamically inequivalent configurations without mechanical rotation or magnetic-field reorientation. Squeezing one of the two magnon modes produces a high-contrast nonreciprocal response ($C_E, C_{\mathcal R} \simeq 1$) in selected stable windows, while choosing whether $m_1$ or $m_2$ is squeezed changes the dominant optomagnonic entanglement channel. The results show that the contrast depends on the squeezing strength, squeezing phase, magnomechanical coupling, and bath temperature, with phonon-mediated correlations limited by the thermal occupation of the MHz mechanical modes. Moreover, the genuine quantum nature of the squeezing is confirmed by Wigner function reconstructions at opposite phases. Within the assumptions of the linearized Gaussian treatment and for experimentally motivated parameters, these results establish a feasible theoretical route toward phase-switchable entanglement in an optomagnomechanical ring cavity, with practical implementation requiring cryogenic precooling, sideband cooling of the mechanical modes, and accurate control of the squeezing phase, all within reach of current experimental platforms. This positions magnon squeezing as a versatile resource for chiral quantum networks and on-demand entanglement routing in hybrid magnonic architectures.

\section*{Acknowledgment}
Z.I. acknowledges the UM6P Vanguard Center for hospitality and support during the completion of this work. 
A.E.A. completed part of this work during a research visit to the Laboratoire de Physique et Modélisation des Milieux Condensés (LPMMC), CNRS, in Grenoble, France. He extends his sincere gratitude to the CNRS - Fédération de Recherche QuantAlps, Comité de Direction QuantAlps, for their financial support and for fostering a stimulating and friendly research environment. The authors are also grateful to the referee for their constructive feedback, which significantly enhanced the quality of this manuscript.

\section*{Appendix A}
\label{appendix:A}

\renewcommand{\theequation}{A\arabic{equation}}
\setcounter{equation}{0}

In this Appendix, we provide the detailed derivation of the rotating-frame Hamiltonian used in the main text. For simplicity, we set $\hbar=1$ throughout the following calculations. The total Hamiltonian is given by
\begin{eqnarray}
	\label{A1}
	\tilde{\mathbf{H}}&=&\omega_c c^\dagger c +\sum_{j=1,2}\omega_{m_j} m_j^\dagger m_j+\frac{\omega_{b_j}}{2}\big({q_j^2+p_j^2}\big)\nonumber\\
	&+&\sum_{j=1,2} \Big[ \bar{g}_{0j}c^\dagger c\,{q_j}+g_{m_j}m_j^\dagger m_j\,{q_j} + i\Big( \frac{\Gamma_j}{2}e^{i\theta_j}e^{-i2\omega_{0}t}m^{\dagger2}_j\nonumber
	\\&+&\Omega_j m_j^\dagger e^{-i\omega_{0}t}\Big)-\text{H.c.}\Big] +(\Upsilon c^\dagger e^{-i\omega_{L}t}-\text{H.c.}).
\end{eqnarray}
By applying the unitary transformation $\mathcal{U}=\exp\big[i(\omega_L c^\dagger c + \sum_{j=1,2} \omega_0 m_j^\dagger m_j)t\big]$ to the total Hamiltonian in Eq.~(\ref{A1}), 
the system operators transform as 
\begin{eqnarray}
	c \to c \, e^{-i\omega_L t} &~~,~~& c^\dagger \to c^\dagger \, e^{i\omega_L t},\\
	m_j \to m_j \, e^{-i\omega_0 t} &~~,~~& m_j^\dagger \to m_j^\dagger \, e^{i\omega_0 t},\\
	m_j^2 \to m_j^2 \, e^{-i2\omega_0 t} &~~,~~& m_j^{\dagger2} \to m_j^{\dagger2} \, e^{i2\omega_0 t},
\end{eqnarray}
and we obtain the following interaction-picture Hamiltonian:
\begin{eqnarray}
	\label{X1}
	{\mathbf{H}}&=&\mathcal{U}\tilde{\mathbf{H}}\mathcal{U}^\dagger-\text{i}~\mathcal{U}\partial_t \mathcal{U}^\dagger\nonumber\\
	&=&(\omega_c-\omega_L) c^\dagger c +\sum_{j=1,2} (\omega_{m_j}-\omega_0) m_j^\dagger m_j+ \frac{\omega_{b_j}}{2}(q_j^2+p_j^2) \nonumber\\
	&+& \sum_{j=1,2}\left[\bar{g}_{0j}c^\dagger c q_j+g_{m_j}m_j^\dagger m_jq_j + i (\frac{\Gamma_j}{2}e^{i\theta_j} m^{\dagger2}_j -\text{H.C} )\right]\nonumber\\
	&+&	i\left[(\Upsilon c^\dagger +\sum_{j=1,2} \Omega_j m_j^\dagger)-\text{H.C}\right],
\end{eqnarray}
However, the quantum Langevin equations including dissipation and noise turn out to be:
\begin{eqnarray}
	\label{X2}
	\dot{c}&=& -i\Delta_c c-\kappa_c c-i\sum_{j=1,2}{\bar{g}_{0j}} cq_j+ \Upsilon + \sqrt{2\kappa_c} c^{in},\nonumber\\
	\dot{m}_j &=& -i\Delta_{m_j}m_j-\kappa_{m_j}m_j-ig_{m_j}m_jq_j+\Omega_j\nonumber\\&+& \Gamma_j e^{i\theta_j} m^{\dagger}_j + \sqrt{2\kappa_{m_j}} m^{in}_j,\nonumber\\
	\dot{q}_j&=&\omega_{b_j}p_j,\nonumber\\
	\dot{p}_j&=&-\omega_{b_j}q_j-g_{m_j}m_j^\dagger m_j-\bar{g}_{0j}c^\dagger c-\gamma_{b_j} p_j+\xi_j,
\end{eqnarray}
where $\Delta_c=\omega_c-\omega_L$ and $\Delta_{m_j}=\omega_{m_j}-\omega_0$. The next step is to linearize the above differential system around steady-state values by writing each operator as $\mathcal{O}=\mathcal{O}_s+\tilde{\mathcal{O}}$, where $\mathcal{O}_s$ is the steady-state value and $\tilde{\mathcal{O}}$ represents small quantum fluctuations. In fact, by neglecting second-order terms yields to: 
\begin{eqnarray}
	\label{X3}
	\dot{\tilde{c}}&=& -i\tilde{\Delta}_c\tilde{c}-\kappa_c\tilde{c}-i\sum_{j=1,2}\bar{G}_0\tilde{q}_j +\sqrt{2\kappa_c} \tilde{c}^{in},\nonumber\\
	\dot{\tilde{m}}_j &=& -i\bar{\Delta}_{m_j}\tilde{m}_j-\kappa_{m_j}\tilde{m}_j-iG_{m_j}\tilde{q}_j + \Gamma_j e^{i\theta_j} \tilde{m}^{\dagger}_j\nonumber\\ &+& \sqrt{2\kappa_{m_j}} \tilde{m}^{in}_j,\nonumber\\
	\dot{\tilde{q}}_j&=&\omega_{b_j}\tilde{p}_j,\nonumber\\
	\dot{\tilde{p}}_j&=&-\omega_{b_j}\tilde{q}_j+\frac{\bar{G}_0}{i\sqrt{2}}(\tilde{c}^\dagger- \tilde{c})-\frac{G_{m_j}}{i\sqrt{2}}(\tilde{m}_j^\dagger- \tilde{m}_j)\nonumber\\ &-&\gamma_{b_j} \tilde{p}_j+\xi_j,
\end{eqnarray}
where $\tilde{\Delta}_c=\Delta_c-\sum_{j=1,2}(-1)^j g_0\cos^2(\Psi/2)q_{js}$, $\bar{\Delta}_{m_j}=\Delta_{m_j}+g_{m_j}q_{js}$, $\bar{G}_0=(-1)^j G_0$, and $G_0=i g_0\sqrt{2}\cos^2(\Psi/2)\alpha_s$.
Introducing quadrature operators $\tilde{A}_\varepsilon=(\tilde{\varepsilon}+\tilde{\varepsilon}^\dagger)/\sqrt{2}$ and $\tilde{B}_\varepsilon=(\tilde{\varepsilon}-\tilde{\varepsilon}^\dagger)/(i\sqrt{2})$ for $\varepsilon=c,m_j$ gives rise to the linearized quantum Langevin equations as follows:
\begin{eqnarray}
	\label{X5}
	\dot{\tilde{A}}_c&=&\tilde{\Delta}_c \tilde{B}_c-\kappa_c \tilde{A}_c +\bar{G}_\Sigma\tilde{q}_j+\sqrt{2\kappa_c} \tilde{A}_c^{in},\nonumber\\
	\dot{\tilde{B}}_c&=&-\tilde{\Delta}_c\tilde{A}_c-\kappa_c\tilde{B}_c+\sqrt{2\kappa_c} \tilde{B}_c^{in},\nonumber\\
	\dot{\tilde{A}}_{m_j}&=& \bar{\Delta}_{m_j} \tilde{B}_{m_j}-\kappa_{m_j} \tilde{A}_{m_j} + {\Delta_{\theta_j}} \tilde{B}_{m_j} + {\kappa_{\theta_j}} \tilde{A}_{m_j} \nonumber\\
	&-&G_{m_j}\tilde{q}_j+\sqrt{2\kappa_{m_j}} \tilde{A}_{m_j}^{in},\nonumber\\
	\dot{\tilde{B}}_{m_j}&=&-\bar{\Delta}_{m_j}\tilde{A}_{m_j}-\kappa_{m_j}\tilde{B}_{m_j}+ {\Delta_{\theta_j}}  \tilde{A}_{m_j} - {\kappa_{\theta_j}} \tilde{B}_{m_j} \nonumber\\
	&+&\sqrt{2\kappa_{m_j}} \tilde{B}_{m_j}^{in},\nonumber\\
	\dot{\tilde{q}}_j&=&\omega_{b_j}\tilde{p}_j,\nonumber\\	
	\dot{\tilde{p}}_j&=&-\omega_{b_j}\tilde{q}_j-\bar{G}_0\tilde{B}_c+G_{m_j}\tilde{B}_{m_j}-\gamma_{b_j}\tilde{p}_j+\xi_j,
\end{eqnarray}
where $\bar{G}_\Sigma=\sum_j(-1)^j G_0$, $\kappa_{\theta_j} = \Gamma_j \cos\theta_j$, and $\Delta_{\theta_j} = \Gamma_j \sin\theta_j$. These equations can be written compactly as $\dot{\digamma}(t) = \chi\digamma(t) + \mathcal{N}(t)$, where {$\digamma(t)$} contains all quadrature fluctuations, $\mathcal{N}(t)$ is the noise vector, and $\chi$ is the drift matrix given in Eq.~\ref{5} in the main text.

\section*{Appendix B: Stability Analysis}
\label{appendix:B}
\renewcommand{\theequation}{B\arabic{equation}}
\setcounter{equation}{0} 

\begin{figure}[t]
	\begin{center}
		\subfigure{\includegraphics[scale=0.16]{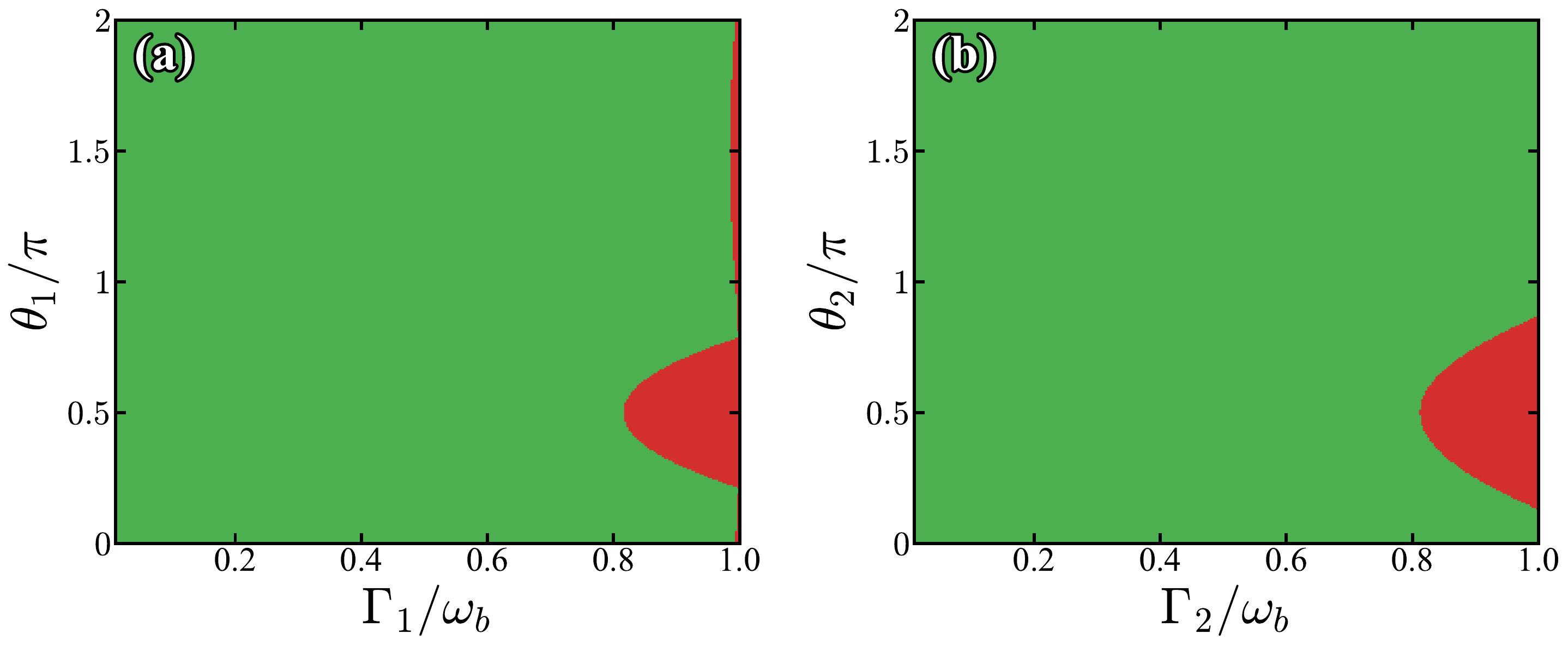}}
	\end{center}
	\vspace{-1.5em} 
	\captionsetup{justification=RaggedRight, singlelinecheck=false}
	\caption{{\small  Stability functions versus $\theta_j$ and $\Gamma_{j}$ in both (a) scenario I, and (b) scenario II. Green (red) regions indicate stable (unstable) parameter regimes.}}
	\label{fig:7}
\end{figure}

The stability of the proposed ring-cavity optomagnomechanical system is verified using the Routh--Hurwitz criterion~\cite{E8}. The linearized dynamics is governed by the drift matrix $\chi$ given in Eq.~\ref{5}. Note that the stability of our system requires that all eigenvalues $\lambda$ of $\chi$ should have negative real parts, which are determined by the characteristic equation $\det(\chi - \lambda I) = 0$. This yields a tenth-order polynomial equation as follows: 
\begin{equation}
	a_0 \lambda^{10} + a_1 \lambda^9 + \cdots + a_9 \lambda + a_{10} = 0.
	\label{eq:charpoly}
\end{equation}
Rather than solving this polynomial equation directly, we apply the Routh--Hurwitz criterion, which provides algebraic stability conditions based solely on the polynomial coefficients ${a_k}$. 
We construct a sequence of Hurwitz matrices $T_k$ of dimension $(k \times k)$ for $k = 1, 2, \ldots, 10$, with elements $t_{ln}$ defined by
\begin{equation}
	t_{ln} = 
	\begin{cases}
		a_{2l-n}, & \text{if }   0 \leq 2l - n \leq 10 \text{ with }  1 \leq l, n \leq k, \\
		0, & \text{otherwise}.
	\end{cases}
	\label{eq:Eq02}
\end{equation}
Therefore, the system is stable if and only if all Hurwitz determinants are positive, i.e., $\det(T_k) > 0$ for all $k = 1, 2, \ldots, 10$ \cite{E8}.
We evaluate these stability conditions numerically across the full parameter space. Throughout all results, unstable regions are excluded and marked by red boundaries, ensuring that all reported entanglement lies within experimentally accessible stable regimes. As shown in Fig.~\ref{fig:7}, $\theta = \pi/2$ corresponds to the narrowest stability window, while phases with both {$\Delta_{\theta_j}$ and $\kappa_{\theta_j}$} active yield broader stable regions.


\begin{thebibliography}{9}
\bibitem{E7} The driving magnetic field is defined as $H_{d_j}=\sqrt{2\mu_0\mathcal{P}_{m_j}/(LWc)}$, where $\mu_0$ and $c$ are the magnetic permeability and velocity of electromagnetic waves in vacuum, $L$ and $W$ are the length and width of the YIG microbridge cross-section, respectively. In addition, $\mathcal{P}_{m_j}= (H_d^2/2\mu_0)LWc$ is the microwave drive power, where $H_d^2/2\mu_0$ is the time-averaged energy density per unit volume.
\end{thebibliography}
\end{document}